\documentclass[referee]{aa}
\usepackage{epsfig}

\def\B{{\em BeppoSAX \/}}
\def\G1{GS\thinspace1843+00}
\def\ergs{erg s$^{-1}$}
\def\cdof{\mbox{$\chi^2_\nu$ }}

\begin{document}

\thesaurus{08.02.3; 08.14.1; 08.16.7; 13.25.5}

\title{\B observation of the transient X--ray pulsar \G1}

\author{S. Piraino \inst{1,2} \and A. Santangelo\inst{1} \and
A. Segreto\inst{1} \and S. Giarrusso\inst{1}  \and G. Cusumano\inst{1}  \and
S. Del Sordo\inst{1} \and  N.R. Robba\inst{2} \and D. Dal Fiume\inst{3}
\and M. Orlandini\inst{3} \and T. Oosterbroek\inst{4}
\and A.N. Parmar\inst{4}}

\authorrunning{S.~Piraino et al.}

\titlerunning{BeppoSAX Observations of \G1}

\offprints{piraino@ifcai.pa.cnr.it}

\institute{{Istituto di Fisica Cosmica ed Applicazioni
dell'Informatica, CNR, Via Ugo La Malfa 153, 90146 Palermo, Italy
} \and {Istituto di Fisica, Universit\`{a} di Palermo, via
Archirafi 36, 90123 Palermo, Italy} \and{Istituto Tecnologie e
Studio Radiazioni Extraterrestri, CNR, Via Gobetti 101, 40129
Bologna, Italy} \and{Astrophysics Division, Space Science
Department of ESA, ESTEC, P.O. Box 299,2200 AG Noodwijk, The
Netherlands}}

\date{Received ; accepted}

\maketitle

\begin{abstract}

We present the results from both the timing and spectroscopic
analysis of the transient X--ray pulsar \G1 observed by the \B
satellite on 1997 April 4, when the source was at a luminosity of
$\sim10^{37}$~\ergs. \G1 shows a very hard spectrum that is well
fitted by an absorbed power law (N$_H$ $\sim 2.3 \times 10^{22}$
cm$^{-2}$) modified by a high energy cut--off above 6~keV.  The
source shows a small pulse amplitude in the whole energy band. The
pulse profile evolves with energy from a double--peaked to
single--peaked shape.  The barycentric pulse period is
$29.477\pm0.001$ s.


\end{abstract}

\section{Introduction}

The transient X--ray pulsar \G1 was discovered on 1988 April 3
during a galactic plane scan observation near the Scutum region by
the {\em Ginga \rm} satellite (\cite{Makino88a};
\cite{Makino88b}).  The Large Area Counter (LAC, \cite{Turner89})
on board {\em Ginga \rm} (\cite{Makino87}) detected at a J2000
position of ${\rm \alpha = 18^{h} 45^{m}\pm1^{m}}$, ${\rm \delta =
0^\circ 55'\pm7'}$, a coherent pulsation with a period of 29.5 s
and an 2--37~keV X--ray intensity of 50 mCrab (\cite{Koyama90a}).
On 1988 April 19 and 20 {\em Ginga \rm} carried out a pointed
observation of \G1, measuring a highly variable X--ray flux on a
wide range of time scales, ranging from 30 to 60 mCrab. In
addition to a coherent oscillation with P$= 29.5056 \pm 0.0002$~s,
an energy--dependent aperiodic variation was found (Koyama et~al.
\cite{Koyama90b}).

On 1997 March 3, the Burst and Transient
Source Experiment ({\em BATSE}) on board CGRO detected a new outburst from
this peculiar source (\cite{Wilson97}). The mean 20--50~keV {\it
rms} pulsed flux was $37\pm2$ mCrab while the mean barycentric pulse
period at an epoch of March~6.0 was P$ = 29.5631 \pm 0.0003$~s. The P
variation during this observation implies a spin--up rate, ${\rm \dot
P}$, of $(-3.65\pm0.11)\times10^{-8}$ s~s$^{-1}$.  The data
confirmed the low pulsed fraction ($\sim$7\%) observed by {\em Ginga}.

Between 1997 February~1 and March~19 the All Sky Monitor (ASM) on
board the Rossi X--ray Timing Explorer ({\em RXTE}) observed the
source to be at a flux level of ${\rm F_{2-10} \sim 15-30 }$~mCrab
(\cite{Takeshima97}).

A pointed observation carried out on 1997 March~5 with the {\em RXTE}
Proportional Counter Array (PCA) detected the source at a 2--60~keV
flux level of 62 mCrab, measuring a barycentric pulse period of
$29.565\pm0.002$~s at an epoch of March 5.1712 UT (\cite{Takeshima97}).

On 1997 April~4 the \B Narrow Field Instruments (NFIs) performed a
pointed observation of \G1 (\cite{Piraino98}). The source flux was
$(2.9\pm0.3)\times10^{-9}$erg cm$^{-2}$s$^{-1}$ in the 0.3--100~keV
energy range.

Using the capability of \B imaging instruments, the
90\% confidence J2000 position of \G1 was constrained to be within a
$30''$ radius circular error region centered on ${\rm \alpha = 18^{h}
45^{m} 34^{s}}$, ${\rm \delta = 0^{\circ} 52\farcm5}$ (\cite{Santangelo97}).
On the same day, the source was also observed by the
{\em ROSAT} High Resolution Imager (HRI) which found a J2000 position
for the source of ${\rm \alpha = 18^{h} 45^{m} 36\fs9}$, ${\rm \delta
= 0^{\circ} 51' 45''}$ (90 \% confidence error radius $10''$,
\cite{Dennerl97}).

In this {\it paper} we present the results of both
a timing and spectroscopic analysis of the \B observation of \G1.

\section{Observation}

The Satellite for X--ray Astronomy \B is described in detail in
Boella et~al. \cite{Boella97a}. It carries a pay--load of four
co--aligned Narrow Field Instruments: the Low Energy Concentrator
Spectrometer (LECS, 0.1--10 keV; \cite{Parmar97}), the Medium
Energy Concentrator Spectrometers (MECS, 1.6--10~keV; Boella
et~al. \cite{Boella97b}), the High Pressure Gas Scintillation
Proportional Counter (HPGSPC, 4--100~keV; \cite{Manzo97}) and the
Phoswich Detector System (PDS, 15--200~keV; \cite{Frontera97}).

The LECS and MECS are imaging instruments
with an angular resolution of $\sim 1.2'$. The HPGSPC and PDS are
collimated instruments with field of views (FOVs) of $1^\circ \times
1^\circ$.

The \B Target Opportunity Observation Program of \G1 started at
1997 April~4 at 02:17 and ended at 15:00~UTC. Good data were
selected from intervals when the instrument configurations were
nominal and the elevation angle, with respect to the Earth Limb,
was greater then $5^\circ$. The total on--source exposure times
were 8.5 ks for the LECS, which is operated only during night
time, 22 ks for the MECS, 9.7 for the HPGSPC and 7.4 ks for the
PDS.

The MECS light curves and spectra have been extracted, following the
standard procedure, from a circular region, centered on the source
with a $4'$ radius, while a $8'$ radius was used for the LECS.

For both LECS and MECS, the background subtraction, was performed using
the background obtained from a long blank sky observations, rescaled
by a correction factor obtained as the ratio of the count rates
extracted from both the blank sky and the \G1 images, in a region of
the detector far from the source location. As far as concerns the
two collimated instrument, HPGSPC and PDS, the background was
subtracted using the standard procedure (\cite{Segreto97}; \cite{Frontera97})
that uses the rocking collimator technique.

\section{Temporal Analysis}

\begin{figure}[h]
\centerline{\epsfig{figure=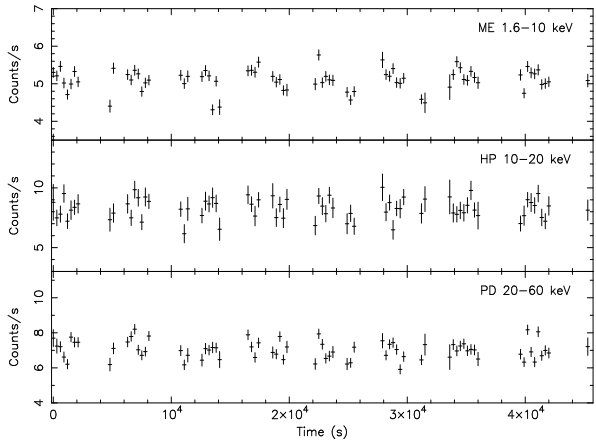,height=7cm,rheight=6.5cm,width=9cm,angle=-90}}
\caption[]{\G1 background subtracted light curves in three energy
ranges.  The gaps are due to South Atlantic Anomaly passages and
Earth occultations} \label{Fig1}
\end{figure}

The arrival times of the photons were first converted to the solar
system barycenter.  In Fig.~\ref{Fig1} we show the background
subtracted X--ray light curves, obtained in three energy ranges,
with 300~s time bin size.  The maximum intensity variation is
$\sim$30\% in the 1.6--10~keV and 20--60~keV energy ranges, and
60\% in 10--20~keV range.

\begin{figure}[h]
\centerline{\epsfig{figure=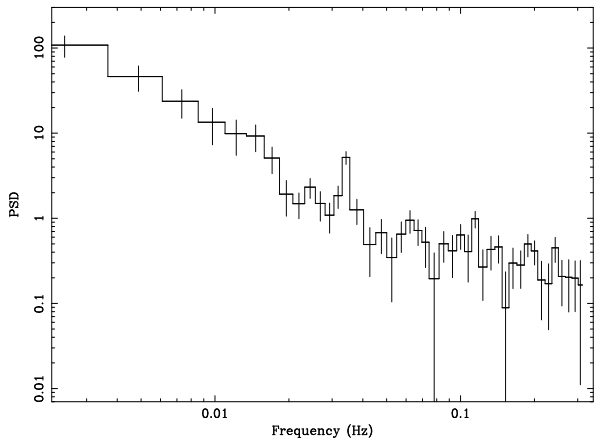,height=7cm,rheight=6.5cm,width=9cm,angle=-90}
}
 \caption[]{\G1 power spectrum in the 1.6--10~keV energy range.
The peak due to the fundamental frequency at 0.0339 Hz and the
aperiodic variability are clearly visible} \label{Fig2}
\end{figure}

\begin{figure}[h]
\centerline{\epsfig{figure=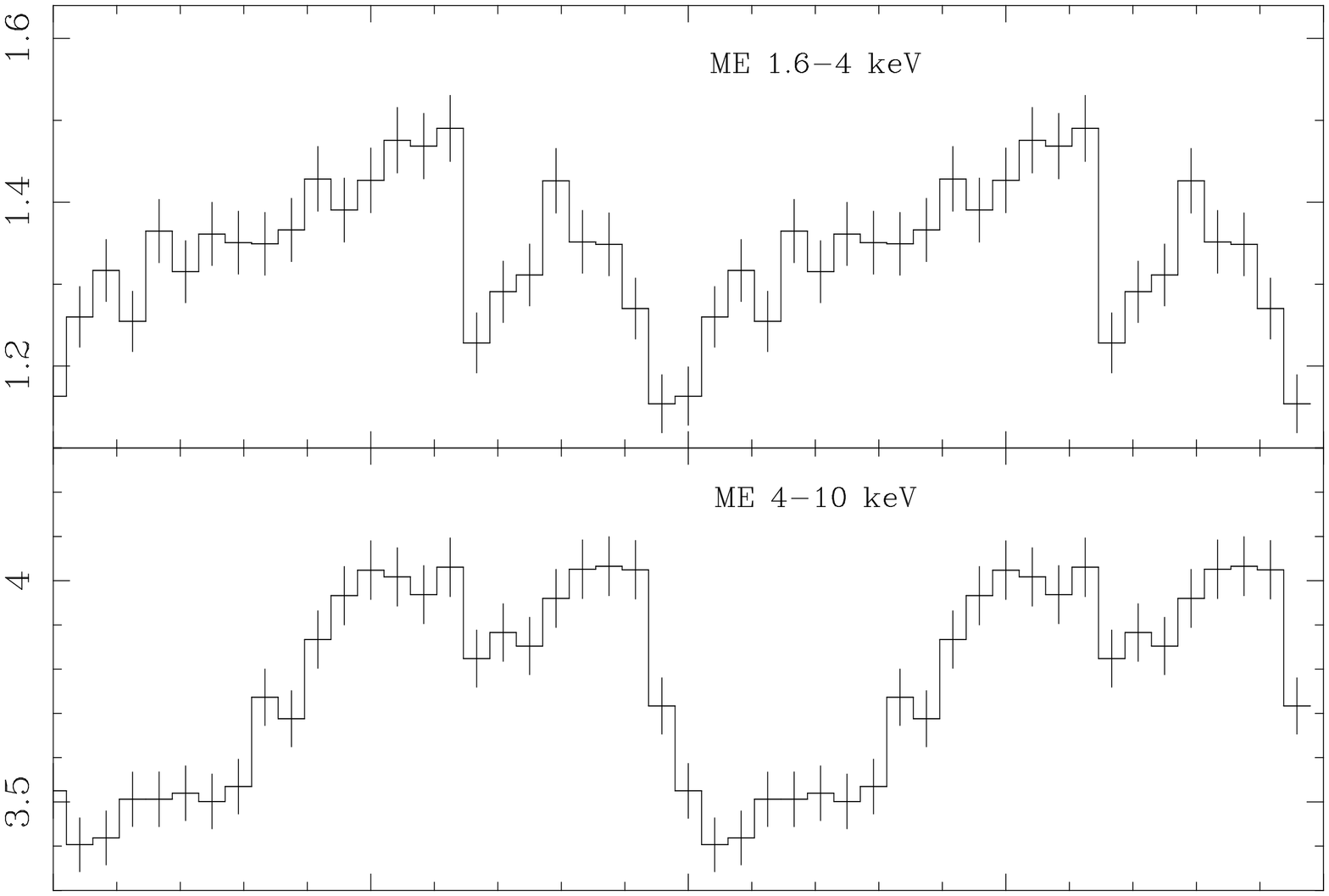,height=6cm,rheight=4.75cm,width=6cm,angle=-90}}
\centerline{\epsfig{figure=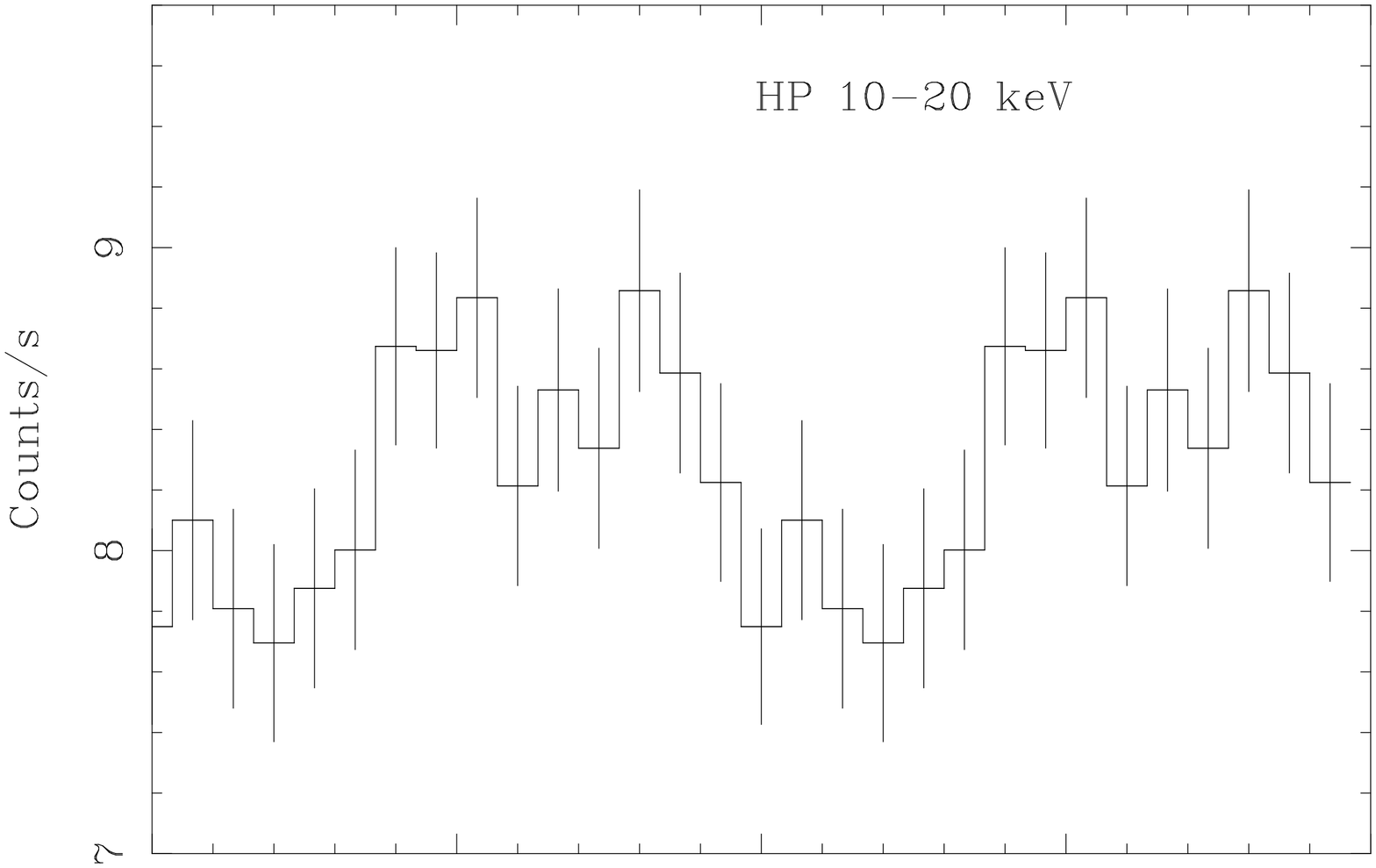,height=3cm,rheight=2.2cm,width=6cm,angle=-90}}
\centerline{\epsfig{figure=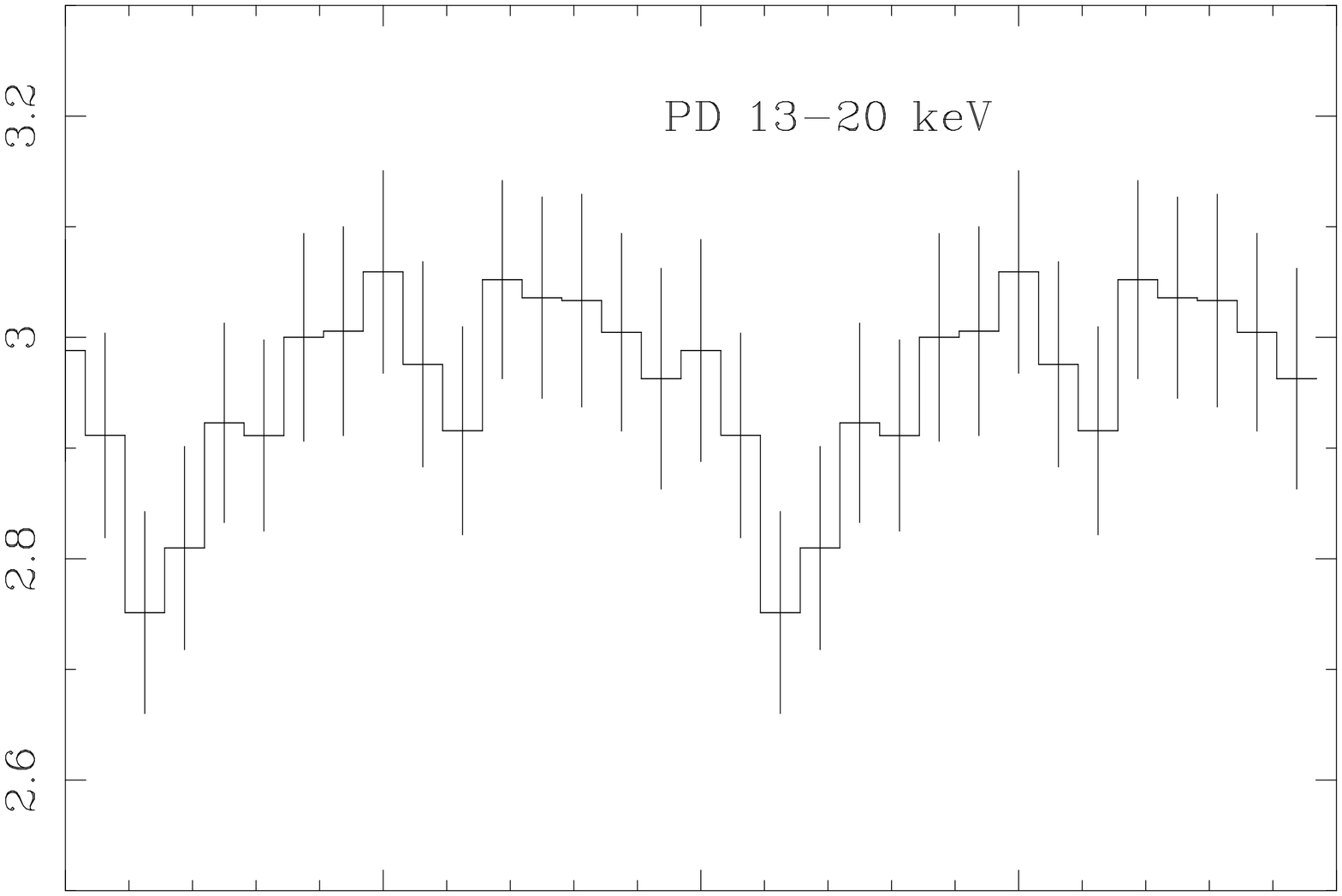,height=3cm,rheight=2.2cm,width=6cm,angle=-90}}
\centerline{\epsfig{figure=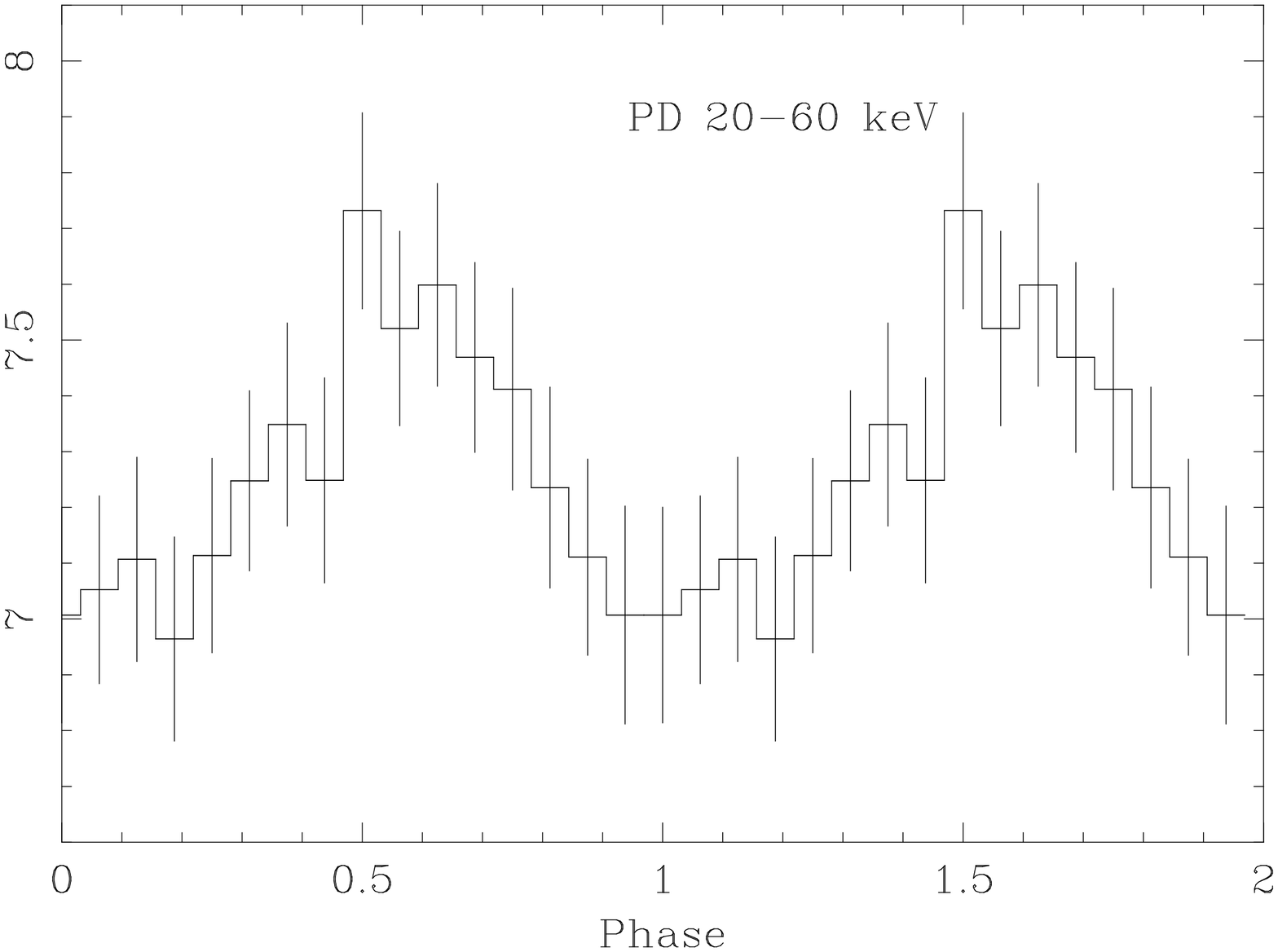,height=3cm,rheight=2.2cm,width=6cm,angle=-90}}
\vspace{0.8cm} \caption[]{\G1 pulse profiles in five energy
ranges. 1$\sigma$ uncertainties are shown} \label{Fig3}
\end{figure}

The (1.6--10 keV) \G1 power spectrum, is shown in Fig.~\ref{Fig2}.
An outstanding peak at 0.0339 Hz is clearly observed.  The \G1
pulse period was obtained with an epoch--folding technique using
barycentric corrected 1.6--10 keV MECS data, while the (1$\sigma$)
uncertainty was determined by fitting the arrival times of sets of
9 averaged profiles, each of 16 phase bins. The best--fit period
is $29.477\pm0.001$.  There is no evidence for any change in
spin--period during the observation with a $\sigma$ upper--limit
of $7.5 \times 10^{-8}$~s~s$^{-1}$.
Using this period value, we folded the light curves in different
energy bands.  The pulse profiles in five energy ranges are shown
in Fig.~\ref{Fig3}.  At lower energies the pulse profile is
clearly asymmetric with a double peak shape, whilst at higher
energies it becomes a simple sinusoid.

\begin{figure}[h]
\centerline{\epsfig{figure=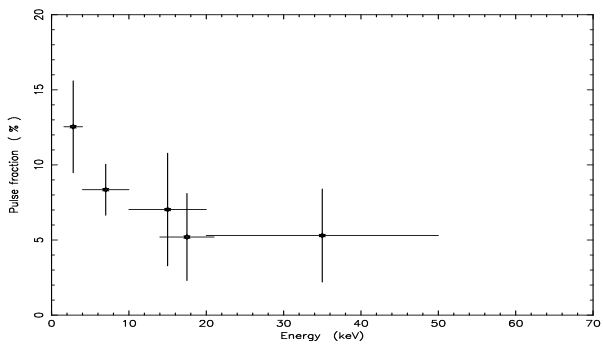,height=5cm,rheight=4.5cm,width=9cm,angle=-90}}
\vspace{0.5cm} \caption[]{The \G1 Pulsed fraction versus energy}
\label{Fig4}
\end{figure}

The variation with energy of the pulsed fraction, defined as the
semi--amplitude of the modulation divided by the average intensity, is
shown in Fig.~\ref{Fig4}.  There is no evidence for an increase in the
fractional periodic variation with energy.

In the \B observation, \G1 may have opposite behavior than during {\it
Ginga} observation, where the pulsed fraction was found to increase
with energy and the pulse profile was clearly single peaked a lower
energies and more structured at higher energies (Koyama et~al. \cite{Koyama90b}).

From {\em BATSE}\footnotemark\footnotetext{{\em BATSE} data are
provided on line by {\em BATSE} pulsar team, at
http://www.batse.msfc.nasa.gov/data/pulsar/sources}, {\em RXTE}
(\cite{Takeshima97}) and \B  data a clear spin--up trend (${\rm
\dot{P}=-3.79\pm0.10\times10^{-8}}$~s~s$^{-1}$ ) over 30 days is
evident (Fig. 5). The mean spin-up timescale, ${\rm P/\dot{P}}$,
is a very rapid 24.6 years. However a difference of $\Delta P$ of
0.01 s is observed
 between the BeppoSAX period and the one expected from the BATSE data extrapolation.
 This could be due to a Doppler effect of orbital motion. Actually,  the change of the pulse
 period, $\Delta P$, due to orbital motion is constrained to be
 \begin{equation}
 \Delta P \leq\frac{P}{c}\frac{ 2 \sin i}{\sqrt{1-e^2}}[\frac{2\pi
 G}{P_{orb}}(M_{NS}+M_{c})]^{\frac{1}{3}}
\end{equation}
where $P_{orb}$ is the orbital period, $e$  the eccentricity, $i$
 the inclination angle, $M_{NS}$  the mass of the neutron star, $M_{c}$
 the  mass of the companion
star, $G$ is the gravitational constant and $c$ the speed of
light. Following  Corbet (1986) relation to estimate the orbital
period (50d) and assuming a mass of 15 $M_\odot$  for the
companion star, typical of  Be star, the upper limit of $\Delta P$
turns out to be $\sim 0.02$ for circular motion.

\begin{figure}[t]
\centerline{\epsfig{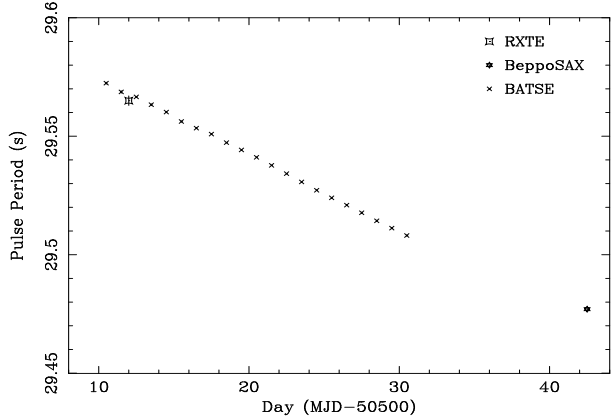}} \vspace{0.8cm} \caption[]{The period  history of the \G1
outburst between early 1997 March to early April obtained from
{\em BATSE}, \B and {\em RXTE}. All the data points are consistent
with a smooth rapid spin--up trend} \label{Fig5}
\end{figure}
To study the aperiodic variability, firstly reported by Koyama et~al. (\cite{Koyama90b})
 the (1.6--10 keV) and (10--37~keV) power spectra were
fitted with a power law. The Poisson white noise was subtracted
from  Leahy normalized power spectra (\cite{Leahy83}). The power
indices, 1.44$\pm$0.17 and 0.9$\pm$0.25 respectively, are
consistent with those found in the {\it Ginga} Observation (Koyama
et~al. \cite{Koyama90b}). The relative amplitude of the aperiodic
variation, calculated dividing the root square of the integrated
PDS over $4\times10^{-3}$ Hz to 10 Hz by an average intensity, is
larger in the lower energy band (18$\%$) than in the higher energy
band (2$\%$).
\section{Spectral Analysis}
Data were selected in the energy ranges 0.1--4 keV, 1.6--10 keV,
8--40 keV and 15--40 keV, respectively for the LECS, MECS, HPGSPC
and PDS, where the instrument responses are well determined and
there are sufficient counts. All spectra have been rebinned to at
least 20 counts for energy channel, in order to ensure the
applicability of $\chi^{2}$ test in the spectral fits.

\begin{figure}[h]
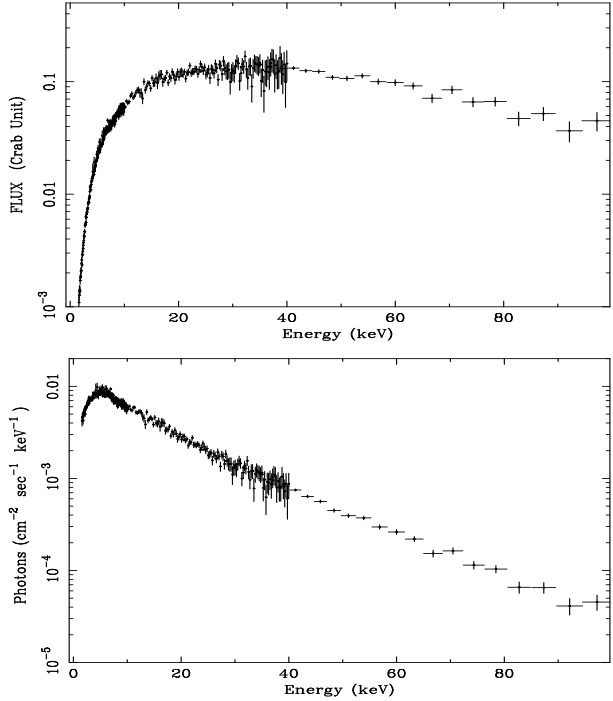


\centerline{\epsfig{figure=7740_f6a.ps,height=4.5cm,width=8cm,angle=-90
}}\vspace{0.2cm}
\centerline{\epsfig{figure=7740_f6b.ps,height=4.5cm,width=8cm,angle=-90
}} \caption[]{ Upper panel: ratio between the \G1 and Crab
spectra. Lower panel: Crab ratio times the functional form of the
Crab spectrum, E$^{-2.1}$ } \label{Fig6}
\end{figure}

\begin{figure}[h]
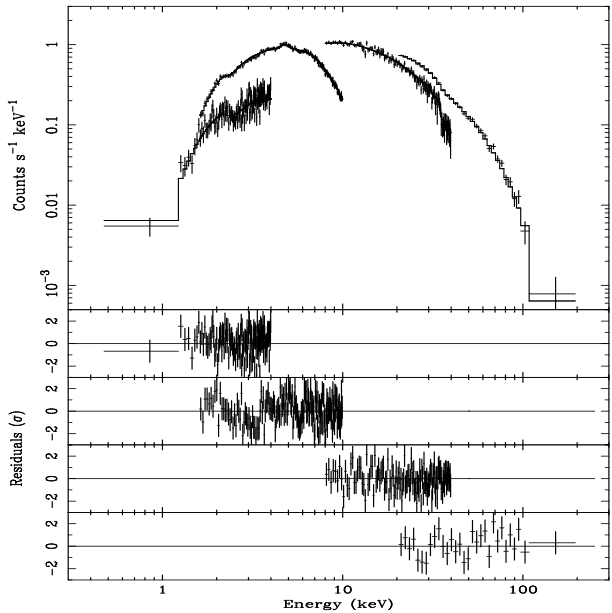

\centerline{\epsfig{figure=7740_f7a.ps,height=4.05cm,rheight=4cm,width=8cm,angle=-90
}}
\centerline{\epsfig{figure=7740_f7b.ps,height=4cm,width=8cm,angle=-90
}} \vspace{0.2cm}\caption[]{Broad band LECS(0.1--4 keV),
MECS(1.6--10 keV), HPGSPC(8--40 keV) and PDS(20--200 keV) X--ray
spectra of \G1 during outburst fitted with model (1).  The lower
panel shows the residuals in terms of sigmas with error bars of
size one.} \label{Fig7}
\end{figure}

Exploiting the \B spectral capability we were able to obtain the
simultaneous broad band spectrum (0.1--200~keV) of GS~1843+00. The
source shows a very hard spectrum strongly absorbed at lower
energies. No deviation from a smooth continuum is observed.  This
can be seen in Fig.~\ref{Fig6} in which the Crab ratio, upper
panel, and the ratio times the functional form of the Crab (a
featureless power--law with $\alpha =2.1 $ in this energy range),
lower panel, are reported.  To extract more physical information
we fitted the phase averaged spectra obtained from the four
co--aligned instruments simultaneously.
The conventional model used to describe the spectrum of X--ray
pulsar (\cite{Pravdo78}; \cite{White83}) is an absorbed power law
with exponential cut--off at higher energies, i.e. a photon
spectrum of the form
\begin{equation}
 {\rm f(E)={A }{E^{-\alpha}}{\exp\{{-N_{H}}{\sigma(E)}-H(E)}\}}
\end{equation}
where E is the photon energy, $\alpha$ is the power--law photon
index, $\rm N_{H}$ is the absorbing column and ${\rm \sigma(E)}$
is the photoelectric absorption cross sections due to cold matter
(\cite{Morrison83}). The high--energy cut--off is modeled by the
function of the form:
\begin{equation}
\rm H(E) = \left \{ \begin{array}{ll} 0 & \mbox{ $E < E_c$} \\ \frac{E
- E_c}{E_f} & \mbox{ $E > E_c$} \end{array} \right.
\end{equation}
where ${\rm E_c}$ is the cut--off energy and ${\rm E_f}$ is the
e--folding energy.\\ Using this model, we obtained a \cdof of 1.08
for 477 degrees of freedom (dof).  The best--fit parameters are
summarized in Table.~\ref{tab:Tab1}.
\begin{table}[h]
\caption{Spectral Parameters for the \G1 broad band fit. All
quoted uncertainties are at 90\% confidence for a single parameter
($\Delta \chi^2 = 2.7$)} \label{tab:Tab1}
\begin{center}
\begin{tabular}{rll}
\hline \noalign{\smallskip} Parameter & Value & Units \\ \hline
\noalign {\smallskip} ${\rm N_{H}}$ & 2.30$ \pm $ 0.13 & $10^{22}$
cm$^{-2}$ \\ $\alpha$ & 0.34$ \pm $ 0.04 & \\ ${\rm E_c}$ & 5.95$
\pm $ 0.45 & (keV) \\ ${\rm E_f}$ & 18.4$ \pm $ 0.6 & (keV) \\
\cdof\ (dof) & 1.08(477) & \\ \noalign{\smallskip} \hline
\end{tabular}
\end{center}
\end{table}
The spectrum together with the best--fit model are shown in the
upper panel of Fig.~\ref{Fig7}.
Fit residuals in terms of $\sigma$ are reported in the lower panel
and its show no clear evidence of any absorption or emission
features.  Normalization factors, between the instruments, were
left free in the fits. Setting the MECS as reference, the relative
normalizations are 0.83 for the LECS, 1.02 for the HPGSPC and 0.79
for the PDS. These values are in good agreement with the ones
obtained from the intercalibration analysis of the four Narrow
Field Instruments (\cite{Fiore99}). The inclusion in the model of
a gaussian line gives a marginal improvement in fit quality (at
less then 90$\%$ confidence level) for a fluorescent K$_\alpha$
line at 6.4 keV with a flux level of $(2.4\pm 1.0) \times10^{-4}$
photons~cm$^{-2}$ s$^{-1}$.

\section{Discussion}
After its discovery in 1988, \G1 was detected again in 1997 March
as a bright (0.3--100~keV) $\sim 2.9\times10^{-9}$ erg
cm$^{-2}$s$^{-1}$ X--ray source.  Due to the spatial capabilities
of the \B imaging instruments an improved position was obtained.
The \B position is within the {\it Ginga} (Koyama et~al.
\cite{Koyama90b}) and RXTE (\cite{Chakrabarty97}) error boxes, and
is also consistent with that measured by the {\it ROSAT} HRI
(\cite{Dennerl97}). Accurate measurement of the position of the
source is important in order to carry out a systematic search for
the still unidentified optical counterpart. Pulsations with a
period $P=29.477\pm 0.001$~s together a mean pulse period change
${\rm \dot{P}/P =-4.1\times10^{-2}}$ yr$^{-1}$, which is in good
agreement with the one measured by {\it Ginga} were found. Koyama
et~al. (\cite{Koyama89}) suggested that such a high spin up rate
could be due, at least partly, to an orbital Doppler motion. Pulse
period variations observed in the 30 days monitoring obtained by
combining data from {\em BATSE}, {\em RXTE--PCA} and \B, confirmed
the presence of a high intrinsic spin--up rate. Moreover, assuming
a Be transient system having an orbital period between 50 and 60
days, inferred from the pulse--orbital periods relation of Corbet
(\cite{Corbet86}), a possible  Doppler effect may be overlapped to
this intrinsic spin--up rate.

The source spectrum, which is well described by an absorbed power
law with high energy cut--off, is typical of accreting X--ray
pulsars.  The very high absorption, ${\rm N_{H} = 2.3 \times
10^{22} cm^{-2}}$ is consistent with that reported by Koyama
et~al. (\cite{Koyama90b}). The hypothesis that the absorption is
mainly interstellar rather then circumstellar (\cite{Koyama90a})
is supported by the marginal detection of a fluorescent K$_\alpha$
iron line in the source spectrum.

Assuming a distance of 10 kpc (\cite{Koyama90a}; \cite{Hayakawa77})
the 0.3--100~keV luminosity is $\sim$$3\times10^{37}$~erg~ s$^{-1}$.

It is unclear if cyclotron resonance scattering features are
present in the hard X--ray spectrum of the source.  Koyama et~al.
(\cite{Koyama90b}) suggested that the cut--off in the spectrum
observed at $\sim$18~ keV could be related to a very intense
magnetic field typical of this class of source. Moreover, Mihara
(\cite{Mihara}), fitting the phase resolved spectrum with an
absorption--like feature at $\sim$20~keV, classified \G1 as a
possible cyclotron source.  Although the spectrum is observed with
good statistics up to $\sim$100~keV, no evidence of any cyclotron
feature is observed in the \B pulse phase averaged spectrum of
\G1.  Also the "crab--ratio technique"(\cite{DalFiume98}),
successfully exploited in detecting Resonance Cyclotron Features
(RCFs) in other X--ray pulsars, does not display any sign of
cyclotron features. Moreover no evidence of cyclotron absorption
features was found in the phase resolved spectra below 100 keV.
However, we found an upper limit on the depth of 0.15 for the
possible 20 keV feature. This value is compatible with that found
by Mihara (\cite{Mihara}).

 Manchanda (\cite{Manchanda99}), using data from the LASE experiment, a
balloon--born large area scintillation counter, recently suggested
the possibility of an absorption feature around 100 keV or an
emission at 140 keV. Unfortunately, statistics of \B spectra is
quite low at that energy and a much deeper analysis, which is
underway, is required.

There are $\sim$80 known accreting X--ray pulsars (see
\cite{Bildsten97} for a recent review).  Until recently only the
relatively bright nearby pulsars were visible due to the limited
sensitivity of previous detectors.  This is changing with the
discovery by {\em ASCA}, {\em ROSAT}, \B and {\em RXTE} of a
population of faint, absorbed pulsars (e.g., \cite{Angelini98};
\cite{Kinugasa98}, \cite{Torii98}).  The search for faint pulsars
is one of the main scientific objectives of the {\em ASCA}
galactic plane survey (e.g., \cite{Sugizaki97}; \cite{Torii98}).

\begin{acknowledgements}  We wish to thank the referee, Toshiaki Takeshima, for
the precise and detailed comments which improved the quality of
the paper. We thank, also, the \B\ Scientific Data Center staff
for their support during the observation and data analysis. SP
acknowledges support from CNR PhD grant. This research has been
partially funded by the Italian Space Agency.
\end{acknowledgements}

\end{document}